\documentclass[referee]{raa}           

\usepackage{graphicx,times}
\usepackage{natbib}
\usepackage{amssymb,amsmath}
\bibpunct{(}{)}{;}{a}{}{,}
\usepackage{url}
\usepackage{color}
\usepackage{rotating}

\usepackage[pagebackref=true]{hyperref}

\newcommand{\Mob}{M^{\rm ob}}
\newcommand{\zob}{z^{\rm ob}}

\newcommand{\avg}[1]{\langle #1 \rangle}

\def\gsim{\;\rlap{\lower 2.5pt
 \hbox{$\sim$}}\raise 1.5pt\hbox{$>$}\;}
\def\lsim{\;\rlap{\lower 2.5pt
 \hbox{$\sim$}}\raise 1.5pt\hbox{$<$}\;}

\begin{document}

   \title{Constraints on dark energy from the CSST galaxy clusters}

 \volnopage{ {\bf 20XX} Vol.\ {\bf X} No. {\bf XX}, 000--000}
   \setcounter{page}{1}

   \author{Yufei Zhang 
   \inst{1,2} \and Mingjing Chen\inst{1,2}  \and Zhonglue Wen
      \inst{3,4}
   \and Wenjuan Fang \inst{1,2}
   }

   \institute{ CAS Key Laboratory for Research in Galaxies and Cosmology, Department of Astronomy, University of Science and Technology of China, Hefei, Anhui, 230026, People’s Republic of China; {\it zyfeee@mail.ustc.edu.cn}; {\it mingjing@mail.ustc.edu.cn}; {\it Corresponding author: wjfang@ustc.edu.cn}\\
        \and
             School of Astronomy and Space Science, University of Science and Technology of China, Hefei, Anhui, 230026, People’s Republic of China\\
	\and
National Astronomical Observatories, Chinese Academy of Sciences, 20A Datun Road, Chaoyang District, Beijing 100101, People’s Republic of China; {\it Corresponding author: zhonglue@nao.cas.cn}\\
\and 
CAS Key Laboratory of FAST, NAOC, Chinese Academy of Sciences, Beijing 100101, People’s Republic of China\\
\vs \no
   {\small Received 20XX Month Day; accepted 20XX Month Day}
}

\abstract{We study the potential of the galaxy cluster sample expected from the China Space Station Telescope (CSST) survey to constrain dark energy properties. By modelling the distribution of observed cluster mass for a given true mass to be log-normal and adopting a selection threshold in the observed mass $M_{200m} \geq 0.836 \times 10^{14} h^{-1}M_{\odot}$, we find about $4.1 \times 10^{5}$ clusters in the redshift range $0 \leq z \leq 1.5$ can be detected by the CSST. We construct the Fisher matrix for the cluster number counts from CSST, and forecast constraints on dark energy parameters for models with constant ($w_0$CDM) and time dependent ($w_0w_a$CDM) equation of state. In the self-calibration scheme, the dark energy equation of state parameter $w_0$ of $w_0$CDM model can be constrained to $\Delta w_0 = 0.036$. If $w_a$ is added as a free parameter, we obtain $\Delta w_0 = 0.077$ and $\Delta w_a = 0.39$ for the $w_0w_a$CDM model, with a Figure of Merit for  ($w_0,w_a$) to be 68.99. Should we had perfect knowledge of the observable-mass scaling relation (``known SR" scheme), we would obtain $\Delta w_0 = 0.012$ for $w_0$CDM model, $\Delta w_0 = 0.062$ and $\Delta w_a = 0.24$ for $w_0w_a$CDM model. The dark energy Figure of Merit of  ($w_0,w_a$) increases to 343.25. This indicates again the importance of calibrating the observable-mass scaling relation for optically selected galaxy clusters. By extending the maximum redshift of the clusters from $z_{max} \sim 1.5$ to $z_{max} \sim 2$, the dark energy Figure of Merit for ($w_0,w_a$) increases to 89.72 (self-calibration scheme) and 610.97 (``known SR" scheme), improved by a factor of $\sim 1.30$ and $\sim 1.78$, respectively. We find that the impact of clusters' redshift uncertainty on the dark energy constraints is negligible as long as the redshift error of clusters is smaller than 0.01, achievable by CSST. We also find that the bias in logarithm mass must be calibrated to be $0.30$ or better to avoid significant dark energy parameter bias.
}

   \authorrunning{Y.-F. Zhang et al. }            
   \titlerunning{Constraints on dark energy from the CSST galaxy clusters}  
   \maketitle

%
\section{Introduction}           
\label{sect:intro_de_cluster_csst}
The concordance $\Lambda$ cold dark matter ($\Lambda$CDM) model has proven to provide an accurate description of the Universe \citep{Planck2018}. Nevertheless, the main constituents in the model, namely dark energy (DE) and dark matter (DM), still need explanations from fundamental physics. Questions such as whether dark energy is indeed the cosmological constant \citep{Zeldovich:1967, Zeldovich:1968}, or whether modified gravity theory rather than DE and DM actually explains the observable Universe \citep{Hu2007PhRvD, Capozziello2011PhR...509..167C}, remain unsolved. These questions provide motivations to look for alternatives to the $\Lambda$CDM model. Future high precision cosmological surveys will constrain various such models and clarify many unsolved fundamental questions.

Dark energy, unlike other known forms of matter or energy, is a component postulated to
cause the late time accelerated expansion of the Universe with a negative pressure \citep{Riess1998, Perlmutter1999}. Multiple observational evidences point to the existence of dark energy. However, to date there are no compelling theoretical explanations yet. The most popular model of DE is the cosmological constant whose equation of state (EoS) is $-1$. There are other models in which the equation of state of DE is not $-1$ or even not constant but time-dependent, for example, quintessence \citep{Ratra1988}, phantom \citep{Caldwell2002}, and quintom \citep{Feng2005}. The accelerated expansion of the universe may even imply that gravity should be described by a modified theory of gravity, rather than the standard theory of General Relativity \citep{Heisenberg2019}. Different models of dark energy leave different signatures in the expansion rate of the Universe and the growth rate of structure. Thus surveys that observe the Universe's supernovae, galaxies, galaxy clusters, etc., can potentially reveal the nature of dark energy \citep[e.g.,][]{zrp+17}.

In the hierarchical structure formation scenario, small density fluctuations generated in the primordial Universe act as the seeds for the formation of the Universe's structure. Overdensities in early Universe grow through gravitational instability and hierarchically form larger and larger structures \citep{Peebles1980,Colberg1999}. Galaxy clusters are the largest virialized objects in the Universe. Searches for galaxy clusters have been carried out for decades from multiwavelength data, such as the millimeter wavebands \citep[e.g.,][]{PlanckSZ2016,deHaan2016,Bleem2020}, the optical wavebands \citep[e.g.,][] {2009ApJS..183..197W,Rozo2010,Oguri2018,CostanziSDSS2019,Abbott2020,Wen2021}, and the X-ray wavebands \citep[e.g.,][]{Vikhlinin2009,Clerc2012,Rapetti2013,Bohringer2017,Pacaud2018}. The abundance and spatial distribution of galaxy clusters are sensitive to Universe's expansion and growth rate and hence underlying cosmological model \citep{Allen2011, Kravtsov2012,Weinberg2013}. Clusters have been used to constrain the matter density parameter $\Omega_m$ and the present day rms of linear density fluctuations within a sphere of radius $8h^{-1}$ Mpc, $\sigma_8$\citep[e.g.,][]{Rozo2007,Rozo2010,Rapetti2013,deHaan2016,CostanziSDSS2019,Abbott2020,DESAbbott2021}. Besides, since massive neutrinos suppress matter fluctuations on small scales, this impact on the growth of structure  manifests itself in  cluster observables, which can be used to constrain neutrino mass \citep{Costanzi2013,Mantz2015,PlanckSZ2016}. Clusters have also been demonstrated to provide tight constraints on dark energy from their abundance \citep{Mantz2010, Rozo2010, Mantz2015, deHaan2016}, spatial clustering \citep{Schuecker2003,AbbottDES2019,DESAbbott2021}, and gas mass fractions  \citep{Allen2008,Mantz2014,Mantz2021}. However, the precision on cosmological parameters derived from cluster observables are affected by both theoretical and observational systematic uncertainties.

Forthcoming large surveys, for example the Vera Rubin Observatory \citep{LSST,LSSTSci}, the Euclid space mission \citep{Euclid2011}, and the China Space Station Telescope (CSST) \citep{Zhan2011CSST,Fanzuhui2018,Cao2018,Gong2019}, have the potential to find a large number of clusters. Specifically, the CSST is a 2-meter space telescope planned to be launched in the early 2020s. It will operate in the same orbit as the China Manned Space Station. The CSST aims at surveying 17,500 $\rm deg^2$ sky area over 10 years of operation. Both photometric imaging and slitless grating spectroscopic observations will be conducted. With the unique combination of a large field of view ($\sim$1 $\rm deg^2$), high-spatial resolution ($\sim$0.15 arcsec), faint magnitude limits, and wide wavelength coverage, CSST has great potential to investigate many fundamental problems, such as properties of dark energy and dark matter, validity of General Relativity on cosmic scales, etc. \citep{Zhan2011CSST}. In particular, CSST will detect a large number of  clusters through photometric imaging, spectroscopic observation, and weak gravitational lensing, thanks to its large sky coverage and wide redshift range, which will be valuable for cosmological studies.

We list the key parameters of the CSST survey in Table~\ref{tab:band}.  There are seven photometric and three spectroscopic bands from near-UV to near-IR, namely, NUV, u, g, r, i, z, and y bands for the photometric
survey, and GU, GV, and GI bands for the spectroscopic survey. The
 CSST photometric survey can reach a 5$\sigma$ magnitude limit of $\sim 26$ AB mag for point sources, while for spectroscopic survey, the magnitude limit can reach $\sim 23$ AB mag. The 4000 \AA \ break, Lyman break and 1.6 $\mu$m bump are distinct features to determine photometric redshifts of galaxies. The photometric redshifts of galaxies can be well determined up to the redshift of $z\sim1.4$, at which the 4000 \AA \ break moves to the y band. At higher redshifts, the Lyman break begins to move into the CSST filters. Considering the relative shallow survey depth in the NUV band, it is expected that the photometric redshifts have a larger bias and uncertainty at $1.4<z<2.5$ (even the presence of NUV,    the photometric redshift is not much improved in the redshift range \citep{rtg+15}). They can be improved with the help of other survey whose bands coverage extending to mid-infrared and near-infrared bands, such as WISE \citep{WISE2010} and Euclid \citep{Euclid2011,Sartoris2016}.

\begin{table*}
\centering
\caption{Key parameters of the CSST photometric imaging survey and spectroscopic survey \citep{Zhanhu2021}.}
\label{tab:band}
\begin{tabular}{ c  c  c  c  c  c  c  c c c }
\hline\hline
{surveys} &{area($\rm deg^2$)}&{exposure time(s)}&\multicolumn{7}{c}{magnitude limit(point source, 5$\sigma$, AB magnitude)}\\
\hline
&&& NUV & u & g & r & i & z & y \\
photometric& 17,500 & 150 $\times$ 2 & 25.4 & 25.4& 26.3 &26.0 & 25.9 & 25.2 &24.4\\
       & 400 & 250 $\times$ 8 & 26.7 & 26.7 & 27.5 & 27.2 & 27.0 & 26.4  &25.7\\
\hline
&&& &GU &&GV  & &GI \\
spectroscopic& 17,500 & 150 $\times$ 4 && 23.2 && 23.4 && 23.2&\\
                    & 400 & 250 $\times$ 16 &  & 24.4 & & 24.5 &  & 24.3 & \\
\hline\hline

 \end{tabular}
\end{table*}

In this paper, we explore the power of the CSST cluster sample in constraining dark energy parameters. We consider two dark energy models. The first one is the model in which the equation of state parameter $w$ of dark energy is allowed to deviate from $-1$ in a time independent fashion ($w_0$CDM). In the second model, the equation of state of dark energy is varying with time ($w_0w_a$CDM), with the phenomenological parameterization $w(a) = w_0 + (1-a)\,w_a$, where $a$ is the scale factor of the Universe \citep{Chevallier01, Linder03}. We estimate the abundance of galaxy clusters expected from CSST and forecast its constraints on cosmological parameters by using Fisher matrix technique. To calculate galaxy cluster number counts, we first compute the halo mass function by adopting the fitting function from \cite{Tinker2008}. Then the number counts in an observed mass bin can be computed once the probability to assign an observed mass to a cluster true mass is given. Finally, we can evaluate cluster number counts as a function of the  estimated mass and redshift. By combining the number of clusters in bins of 
 estimated mass and redshift, we construct the Fisher matrix and then derive the forecasted constraints on cosmological parameters. 

This paper is organized as follows. In Section \ref{sec:theory_de_cluster_csst}, we detail our estimation for the galaxy cluster abundance expected for the CSST, and present the Fisher matrix we use to forecast parameter constraints. In Section \ref{sec:result_de_cluster_csst},  we give our results and discuss the effects of several systematics. Finally, we conclude in Section \ref{sec:conclusion_de_cluster_csst}.

\section{Calculational methods}
\label{sec:theory_de_cluster_csst}

\subsection{Mass estimation for CSST clusters}

Cluster mass is of fundamental importance for studies on cluster properties and cluster cosmology. Regardless of how clusters are detected at different wavelengths, the main concern is that halo mass is not directly observable, so we have to employ a suitable observable quantity that scales with mass.  In the case of  optical surveys, a commonly used mass proxy is the optical richness $\lambda$, which corresponds to the count or the total luminosity of member galaxies above some luminosity threshold in a given cluster \citep{Rozo2009,wzl2012}. The calibration of the relation between the richness and halo mass for optically selected clusters can be performed through cluster number counts, clustering and stacked weak-lensing measurements \citep[e.g.,][]{Murata2017,MurataHSC,Chiu1,Chiu2,CostanziSDSS2019,Abbott2020,Wen2021}.

For the CSST survey, each galaxy cluster is identified with an optical richness estimated. Mean mass of clusters can be measured directly through
weak lensing for a sample of stacked clusters within a give richness and redshift bin. Then, one can get an ``accurate" richness-mass scaling relation and its evolution with redshift. In the regime where weak lensing method is not applicable, e.g., very high redshifts, cluster mass can be estimated according to the derived richness-mass relation.

\subsection{Calculation for cosmological constraints}

A fundamental quantity for cluster cosmology is the halo mass function, which is defined as the differential number density of haloes. In this paper, we adopt the halo mass function obtained by \cite{Tinker2008}
\begin{equation}
\frac{dn}{dM} = f(\sigma) \frac{\rho_m}{M} \frac{d\ln\sigma^{-1}}{dM}\,,
\end{equation}
where $f(\sigma)$ is the fitting function given by Eq.(3) in \cite{Tinker2008}, $\rho_m$ is the present matter density of the Universe and $\sigma$ is the rms of linear matter fluctuation within a sphere of radius $R$ that contains mass $M$ given mean density of $\rho_m$. Throughout this paper, we define  cluster mass as $M_{200m} \equiv (4 \pi /3)\, \Delta_{\rm m}\, \rho_{\rm m}(z)\, (R_{\rm 200m})^3$, where $\Delta_{\rm m} \equiv 200$, $R_{\rm 200m}$ is the halo radius within which the mean matter density is 200 times the matter density $\rho_{\rm m}(z)$ of the Universe at  redshift $z$.

To estimate the abundance of galaxy clusters that can be detected from an optical survey, we take into account observational effects such as mass scatter and photometric redshift uncertainty. The average number counts of galaxy clusters expected in a survey with sky coverage $\Omega_{\rm sky}$, within the $m$-th
bin in observed mass $M^\mathrm{ob}$ ($M_{m,min} \le M^\mathrm{ob} \le M_{m,max}$) and $i$-th bin in observed redshift $z^\mathrm{ob}$ ($z_{i,min} \le z^\mathrm{ob} \le z_{i,max}$), can be calculated as
\begin{eqnarray}
N_{m,i} & = & \Omega_{sky} \int_0^\infty dz  \frac{dV}{dz d\Omega}  \int_{z_{i,min}}^{z_{i,max}} d\zob \int_{M_{m,min}}^{M_{m,max}}d\Mob\ \\ \nonumber
 & & \hspace{0.4in} \avg{n|\Mob,z} P(\zob|z)\,. 
 \label{eq:N_mean}
 \end{eqnarray}
In the above expression, $P(\zob|z)$ is the probability distribution function to assign a galaxy cluster at true redshift $z$ to the observed photometric redshift $\zob$, which we model as a Gaussian distribution with expectation value $z$ and scatter $\sigma_z$. 

The comoving space number density of clusters $\avg{n|\Mob,z}$ is related to halo mass function by 
\begin{eqnarray}
\avg{n|\Mob,z} & = & \int_{0}^\infty dM\ \frac{dn}{dM}(M,z) P(\Mob|M,z) \,,
 \end{eqnarray}
where $P(\Mob|M,z)$ is the probability distribution function to assign a galaxy cluster with true mass $M$ and at true redshift $z$ to the observed mass $\Mob$. 

 The survey volume element $dV/(dz d\Omega)$ is given by  
 \begin{eqnarray}
 \frac{dV}{dz d\Omega} & = & cH^{-1}(z)\chi^2(z) \,,
\end{eqnarray}
where $H(z)$ is the Hubble parameter, and $\chi(z)$ is the comoving radial distance to redshift $z$.

The key ingredient in our analysis is the probability distribution function of the observed mass for halos with a given true mass $M$ and redshift $z$, $ P(\Mob|M,z)$. Following \cite{Sartoris2016}, we assume a log-normal distribution function, namely
\begin{eqnarray}
P(\Mob|M,z)~{\rm d} M = \\ \nonumber
 & & \hspace{-1.2in} \frac{1}{\sqrt{2\pi}\sigma_{\ln M}}\exp\left[ -x^2(\Mob,M,z)\right]~{\rm d}\ln M  \,,
\label{eq:p_Mob}
\end{eqnarray}
where  $x(\Mob,M,z)$ is defined as
\begin{equation}
 x(\Mob,M,z) \equiv \frac{1}{\sqrt{2}\sigma_{\ln M}} (\ln \Mob - \ln M_{bias} - \ln M)\,,
  \label{eq:lambda_M}
\end{equation}
where $\ln M_{bias}$ and $\sigma_{\ln M}$ are the bias and scatter of mass estimation in logarithm space, respectively. Following \cite{Sartoris2016}, we parameterize the bias as
\begin{equation}
  \ln M_{bias} = B_{M,0} + \alpha \ln(1+z) \,.
\end{equation}

We assume the following parametrization for the variance of $\ln M$
\begin{equation}
\sigma_{\ln M}^2 =\frac{\sigma_{\ln \lambda}^2 }{B^2} + \kappa \ (1+z)^{2}\,.
    \label{eq:scatter_M}
\end{equation}
 Here the first term comes from the fact that cluster mass commonly scale with optical richness. The variance of $\ln \lambda$ is composed of a constant intrinsic
scatter $D_\lambda$, and a Poisson–like term \citep{Costanzi2021}
\begin{equation}
\sigma_{\ln \lambda}^2 = D_\lambda^2 + \frac{1}{\langle \lambda \rangle} \,,
    \label{eq:scatter_lnlambda}
\end{equation} 
where the term of $1/\langle \lambda \rangle$ is a function of cluster mass and redshift. The
fiducial values of $\sigma_{\ln \lambda}$ and $B$ can be obtained from the scaling relation fitted by \cite{Costanzi2021}
\begin{eqnarray}
\langle \ln \lambda \rangle (M,z) 
&=&  \ln A +  B\ln
\left(\frac{M}{M_{\mathrm{pivot}}}\right) + B_z\ln\left(\frac{1+z}{1+z_{\mathrm{pivot} } } \right) \,.
\label{eq:mean_relation}
\end{eqnarray}
Here, $A$ is the normalization, $B$ is the slope with respect to halo mass and $B_z$ describes the evolution with redshift. The constants $M_{\mathrm{pivot}}$ and $z_{\mathrm{pivot}}$ are pivot halo mass and redshift. We emphasize that Eq.~\ref{eq:mean_relation} is not used for cluster mass estimation, but for the 
scatter of mass estimation.
 
 The second term $\kappa (1+z)^2$ in Eq.~(\ref{eq:scatter_M}) characterizes the projection effects that depend on redshift. The reason for the chosen form of the projection effects is as follows. The photometric redshift error of galaxies usually increases with redshift. 
 In the algorithms of cluster identification, the width of color cut or photometric redshift slice increases with redshift for both the color-based and the photometric redshift-based methods \citep{2009ApJS..183..197W,2014ApJ...785..104R}.
 As is well-known, the dispersion of photometric redshift generally increases with redshift in the form of $1+z$ \citep{Cao2018}. From the perspective of identifying galaxy clusters, in order to obtain the majority of  member galaxies, the width of the photometric redshift slice used to find galaxy clusters also increases as $1+z$, resulting in the corresponding increase of field galaxies projecting into cluster regions as member galaxies. In \cite{2009ApJS..183..197W}, the authors adopted
 a photometric redshift slice of $z\pm0.04(1+z)$ for the SDSS clusters and found a contamination rate of $\sim20\%$ for member galaxies due to the projection effect. The CSST will have a more accurate photometric redshift than the SDSS \citep{Cao2018}, which will enable us to set a narrower photometric redshift slice for selecting member galaxies of clusters.
 In addition, the slitless spectroscopic survey provides accurate redshifts for bright galaxies. It is possible to have a contamination rate of about 10\% at low redshift for massive clusters. Therefore, we assume $\kappa =0.1^2$ as a fiducial choice.
 
We forecast the constraints on cosmological parameters using Fisher matrix technique, which is based on a Gaussian approximation of the likelihood function around the maximum \citep{Tegmark1997ApJ}. The Fisher matrix is defined as
\begin{equation}
 F_{\alpha \beta} \equiv  - \left \langle \frac {\partial^2 \ln  {\cal L}}
{\partial p_{\alpha} \partial p_{\beta}} \right \rangle \, ,
\end{equation}
where $p_{\alpha}$, $p_{\beta}$ represent model parameters,  ${\cal L }$ is the likelihood function, and angle brackets represents ensemble average. 
The marginalized 1$\sigma$ constraint on parameter $p_\alpha$ can then be obtained by
\begin{equation}
  \sigma_{p_\alpha}=\sqrt{(F^{-1})_{\alpha \alpha}} \,.
\end{equation}

In our analysis we choose the galaxy cluster number counts $N_{m,i}$ as observable. The likelihood of $N_{m,i}$ can be modeled as Poisson distribution with the expectation value $\bar{N}_{m,i}$,
\begin{equation}
\ln  {\cal L}(N_{m,i}|\bar{N}_{m,i}) = N_{m,i}\ln \bar{N}_{m,i} - \bar{N}_{m,i} - \ln (N_{m,i}!) \,.
\end{equation}
Thus the Fisher matrix for cluster number counts is 
\begin{equation}
  F_{\alpha \beta}= \sum_{m,i} \frac{\partial \bar{N}_{m,i}}{\partial
    p_\alpha}\frac{\partial \bar{N}_{m,i}}{\partial p_\beta}
  \frac{1}{\bar{N}_{m,i}}\ \,.
\label{eq:fm_nc}
\end{equation}
Here, the sums over $m$ and $i$ run over mass and redshift bins, respectively. 

We adopt the Figure of Merit (FoM) \citep{DETF2006} for DE to quantify the information gains from given probes and experiments, which is inversely proportional to the area encompassed by the ellipse representing the 68.3 percent confidence level
\begin{equation}
{\mathrm{FoM}} = \left[ \det{\mathrm{Cov}}( w_0, w_a)\right]^{-1/2}\ \,,
\label{EQ:FOM_DETF}
\end{equation}
where $\mathrm{Cov}( w_0, w_a)$ is the marginalized covariance matrix for the dark energy equation of state parameters $w_0$ and $w_a$.

In our Fisher matrix analysis, both cosmological parameters and the parameters modeling bias and  scatter in the scaling relation between the observed and true cluster masses are treated as free parameters, and are constrained simultaneously. We assume flat Universe and choose our cosmological parameter set as: $\{h,\Omega_b h^2, \Omega_c h^2, \sigma_8, n_s, w_0, w_a\}$. Fiducial values of these parameters are listed in Table~\ref{tab:fiducial}, which are the best-fit values from Planck 2018 results \citep{Planck2018}. We marginalize over the set of scaling relation parameters $\{ B_{M,0}, \alpha, A, B, B_z, D_\lambda, \kappa \}$ given above,  referred to as nuisance parameters henceforth, whose fiducial values are given in Table~\ref{tab:fiducial}. The values of $B_{M,0}$ and $\alpha$ are chosen according to \cite{Sartoris2016}, while the values of $A, B, B_z $ and $D_\lambda$  are the best-fit values obtained in \cite{Costanzi2021}. The pivot values for halo mass and redshift ($M_{\mathrm{pivot}}$ and $z_{\mathrm{pivot}}$) in Eq.~(\ref{eq:mean_relation}) are taken to be $3 \times 10^{14} h^{-1}M_{\odot}$ and 0.45, respectively, following \cite{Costanzi2021}. 

\begin{table}
\caption{\label{tab:fiducial}The fiducial values of cosmological parameters (upper section) and nuisance parameters (lower section) adopted in this work. }
\begin{center}
\begin{tabular}{ c|c | c}
\hline\hline
 Parameters & Description &Fiducial values \\
\hline\hline
 \multicolumn{3}{c}{Cosmological parameters} \\
 \hline
$h$     & Hubble constant 	   &0.6766  \\
$\Omega_bh^2$      & Baryon density	         &0.02242 \\
$\Omega_ch^2$  &Cold dark matter density  &0.1193   \\
$\sigma_8$      & Normalization of perturbations	        &0.8102 \\
$n_s$   & Spectral index  &0.9665 \\
$w_0$     & DE EoS parameter	         &-1 \\
$w_a$  & DE EoS parameter 	 &0 \\
\hline
\multicolumn{3}{c}{Nuisance parameters}\\
\hline
$B_{M,0}$  &  Constant term of mass bias    &$0$       \\
$\alpha$  &  Coefficient of redshift dependence in mass bias    &$0$       \\
$A$   & $\langle \lambda \rangle$  at pivot mass scale and pivot redshift &$79.8$     \\
$B$  &  Coefficient of mass dependence in $\langle \ln \lambda \rangle$   &$0.93$    \\
$B_z$  &  Coefficient of redshift dependence in $\langle \ln \lambda \rangle$    &$-0.49$       \\
$D_\lambda$  & Intrinsic scatter in $\sigma_{\ln \lambda}$  &$0.217$     \\
$\kappa$  &  Coefficient of projection effect term    &$0.01$       \\
\hline
\end{tabular}
\end{center}
\end{table}

\section{Results and discussions}
\label{sec:result_de_cluster_csst}

In this section, we present the main results of this paper: constraints on the dark energy equation of state parameters from galaxy cluster number counts of CSST, forecasted with the Fisher matrix formalism. Several systematics are also discussed  simultaneously. We consider dark energy models with  constant ($w_0$CDM) and time dependent ($w_0w_a$CDM) dark energy equation of state. We assume that CSST is capable of detecting clusters up to $z \sim 1.5$ by either redshift-based method \citep[e.g.,][]{Wen2021,2021ApJ...909..143Y} or color-based method \citep[e.g.,][]{2014ApJ...785..104R}. We divide the CSST cluster sample into bins both in redshift and halo mass. For redshift, we consider equal-sized bins of width $\Delta z = 0.05$ in the range $0 \leq z \leq 1.5$. For observed halo mass, we take equal-sized logarithmic bins of width $\Delta\ln (\Mob/M_{\odot}) = 0.2$.  We ignore the covariance between different redshift and mass bins.
We assume 17,500 $\rm deg^2$ sky coverage for the CSST optical wide survey. Most CSST galaxies will have photometric redshift uncertainties  of about 0.02 \citep{Gong2019}. Moreover, CSST perform slitless grating spectroscopic survey for bright sources in addition to photometric imaging survey. Taking into account that clusters have multiple bright member galaxies whose spectroscopic redshifts are probably available from CSST slitless or existing spectroscopic surveys, we expect CSST clusters will have an accurate redshift. In this work, we assume CSST clusters' redshift uncertainty to be $\sigma_z/(1+z)$ = 0.001.

\subsection{Mass limit and number counts}

The lower limit of cluster mass corresponds to a given detection threshold in the observed quantity. The limit is adopted to ensure that the cluster sample obtained has a high completeness and also a high purity. According to the analysis based on mock galaxy redshift survey data \citep{2021ApJ...909..143Y}, we adopt a lower mass limit of $M_{200m} \geq 0.836 \times 10^{14} h^{-1}M_{\odot}$ for the CSST cluster sample in order to get a completeness of $\geq 90\%$ and a purity of $\geq 90\%$. This mass limit roughly corresponds to an equivalent mass limit of $M_{500c} \geq 0.7 \times 10^{14}M_{\odot}$\citep{Wen2021}.

\begin{figure}
\begin{center}
{\includegraphics[angle=0, width=0.91\textwidth]{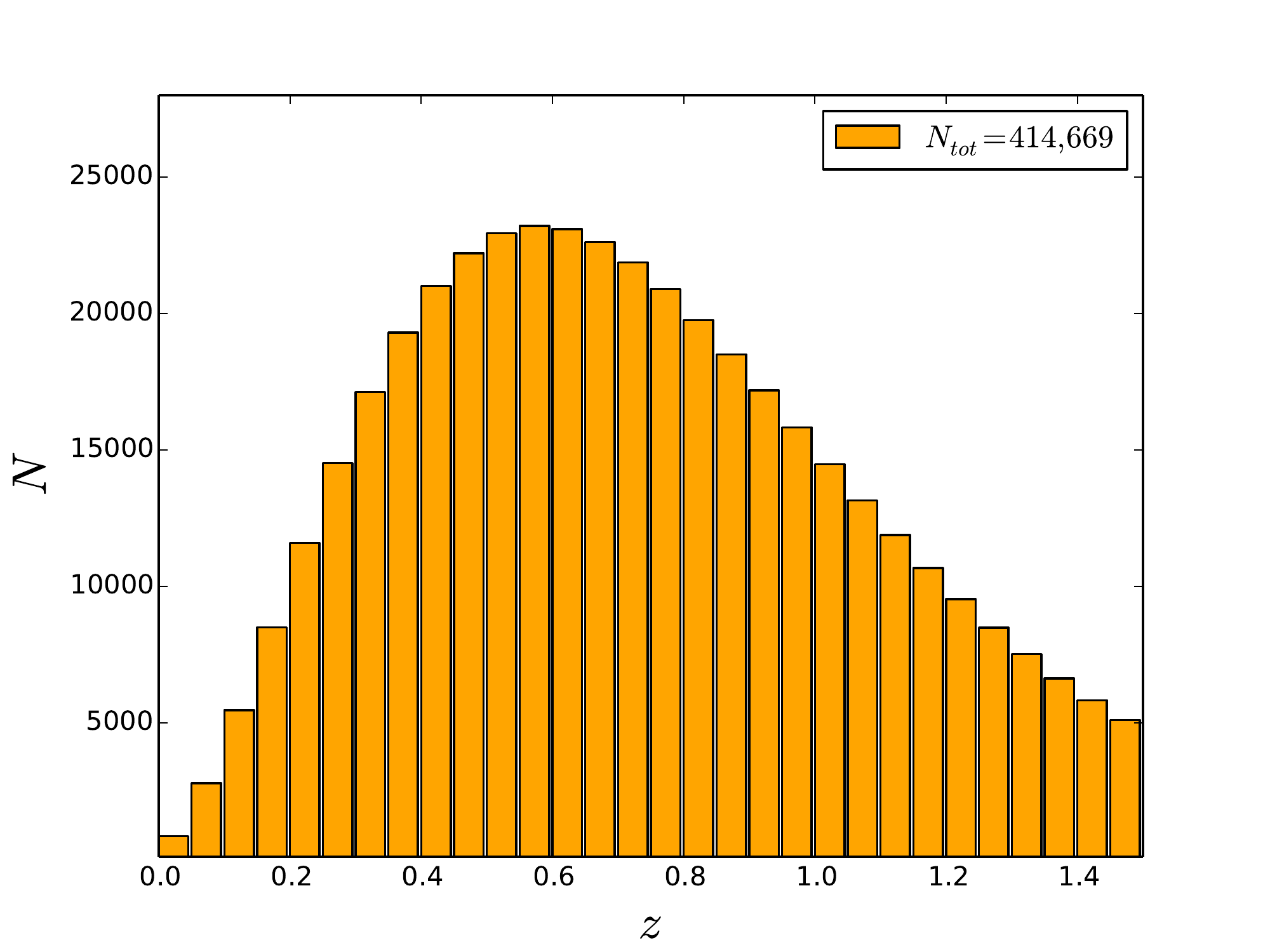}}
 \caption{\label{fig:N_z} Redshift distribution of galaxy clusters expected for CSST, mass threshold is set to $M_{200m} \geq 0.836 \times 10^{14} h^{-1}M_{\odot}$ ($M_{500c} \geq 0.7 \times 10^{14}M_{\odot}$).}
\end{center}
\end{figure}

In Figure~\ref{fig:N_z}, we plot the expected number of clusters that can be detected by CSST as a function of redshift, obtained by adopting the observed halo mass limit $M_{200m} \geq 0.836 \times 10^{14} h^{-1}M_{\odot}$ and assuming the fiducial values of cosmological and nuisance parameters. We find that CSST can detect $\sim 414,669$ clusters in total in the redshift range $0 \leq z \leq 1.5$, with a peak at $ z \sim 0.6$, and there are $\sim 103,069$ clusters at $ z \geq 1.0$. These high redshift clusters are sensitive to the growth rate of perturbations and dark energy properties. A catalogue of uniformly selected high-redshift clusters will be ideal to study structure growth and the underlying cosmological model. 

\subsection{Constraints from cluster number counts}
Since besides cosmological parameters, there are also nuisance parameters which model bias and scatter in the scaling relation between the observed and true cluster masses, we forecast constraints on dark energy equation of state parameters from galaxy cluster number counts of CSST with two schemes: self-calibration scheme in which we take both cosmological parameters and nuisance parameters as parameter entries for the Fisher matrix \citep{2004ApJ...613...41M}, and the ideal case in which the nuisance parameters are fixed, which means that scaling relations are perfectly known in advance (referred to as ``known SR" scheme in the following). 

It is challenging to constrain the cosmological parameters and nuisance parameters simultaneously by using the number counts of galaxy clusters alone. In the following analysis we include the Gaussian priors on the Hubble parameter and the cosmic baryon density from the Planck collaboration \citep{Planck2018} to help break parameter degeneracies. 

The constraints on the cosmological parameters from CSST galaxy cluster number counts are presented in Table~\ref{tab:w_0w_a_results} for the two schemes. In the self-calibration scheme, the dark energy equation of state parameters can be constrained to $\Delta w_0 = 0.036$ for $w_0$CDM model, $\Delta w_0 = 0.077$ and $\Delta w_a = 0.39$ for $w_0w_a$CDM model, corresponding to a dark energy Figure of Merit for ($w_0,w_a$) of 68.99. In the ``known SR" scheme, cluster mass has no bias and the scatter of cluster mass is known. The constraint on $w_0$ is as good as $\Delta w_0 = 0.012$ for $w_0$CDM model, an improvement by a factor of $\sim 3$ compared to the results of the self-calibration scheme, while for $w_0w_a$CDM model, we obtain $\Delta w_0 = 0.062$ and $\Delta w_a = 0.24$. The dark energy Figure of Merit is as high as 343.25, an improvement by a factor of $\sim 5$ compared to the results of the self-calibration scheme. It is apparent that knowledges on the observable-mass scaling relation is essential to get tighter cosmological parameter constraints. The analysis here highlight the importance of the calibration of observable-mass scaling relation for optically selected galaxy clusters in order to obtain tight DE constraints. We postpone a comparison with other optical cluster surveys in Section \ref{sec:comparison_de_cluster_csst}.

It is known that besides dark energy parameters, clusters can also place tight constraints on dark matter related parameters. The 1$\sigma$ uncertainties on $\Omega_ch^2$ and $\sigma_8$ are about $1\% \sim 2\%$ with the self-calibration scheme, while $0.1\% \sim 0.2\%$ with the ``known SR" scheme. The improvements from better knowledge of the observable-mass scaling relation are more pronounced for constraints on $\Omega_ch^2$ and $\sigma_8$ than for $w_0$ and $w_a$. Comparing the constraints from the self-calibration scheme and those from the ``known SR" scheme, the constraint on $\Omega_ch^2$ is improved by a factor of 12.5 for $w_0$CDM model and a factor of 6.4 for $w_0w_a$CDM model, while the constraint on $\sigma_8$ is improved by a factor of 20 for $w_0$CDM model and a factor of 12 for $w_0w_a$CDM model. Thus better calibration of the observable-mass scaling relation is more helpful to tighten the constraints on dark matter related parameters than dark energy parameters. We show  contours of constraints (1 $\sigma$) on cosmological parameters for the $w_0w_a$CDM model in Appendix A.

We point out that in this analysis we assume a possible configuration of CSST. In the following we analyze the impact of two key parameters on our derived constraints, i.e., the maximum redshift and the clusters'  redshift uncertainty. We also compute the requirement for the calibration
of bias in cluster mass.

\begin{table}
\caption{\label{tab:w_0w_a_results} Constraints on the cosmological parameters and the nuisance parameters  from the number counts of CSST galaxy clusters. The column of ``Self-calibration" corresponds to the self-calibration scheme without any priors on nuisance parameters. The column of ``Known SR" refers to the ideal case in which the nuisance parameters are perfectly known. Constraints shown are the marginalized 1$\sigma$ errors. The dark energy FoM is presented in the last row.}
\begin{center}
\begin{tabular}{c|c |c|c |c}
\hline\hline
&\multicolumn{2}{c}{Self-calibration} &\multicolumn{2}{c}{Known SR}\\
 Parameters &$w_0$CDM &$w_0w_a$CDM  &$w_0$CDM &$w_0w_a$CDM\\
\hline\hline
$\Delta \Omega_ch^2$  &0.015   &0.016     &0.0012   &0.0025 \\
$\Delta \sigma_8$      &0.024       &0.025  &0.0012   &0.0021 \\
$\Delta n_s$ &0.077  &0.079  &0.0046   &0.0065 \\
$\Delta w_0$     &0.036         &0.077   &0.012   &0.062\\
$\Delta w_a$  &-   &0.39 &-   &0.24 \\
\hline
$\Delta B_{M,0}$   &0.18 &0.18     &-   &-   \\
$\Delta \alpha$   &0.15 &0.15      &-   &-   \\
$\Delta A$   &31.84 &31.88      &-   &- \\
$\Delta B$  &0.30    &0.31   &-   &-  \\
$\Delta B_z$  &0.55  &0.57  &-   &-      \\
$\Delta D_\lambda$  &0.11 &0.12   &-   &-     \\
$\Delta \kappa$   &0.0064 &0.0067      &-   &-   \\
\hline
FoM  &- &68.99 &-   &343.25 \\
\hline
\end{tabular}
\end{center}
\end{table}

\subsection{Increasing $z_{max}$ of CSST clusters}
 In the above analysis for the CSST clusters, we have adopted a maximum redshift of $z_{max} \sim 1.5$. Higher redshift is potentially achievable with improved cluster selection algorithm, better data quality or joint analysis with the Euclid survey by the European Space Agency \citep{Euclid2011}, which can detect clusters up to redshift as high as $\sim 2$ thanks to the use of near-infrared bands \citep{Sartoris2016}. 

 In this section, we study the impact of including higher redshift clusters on our forecast by increasing the maximum redshift of the CSST cluster sample. Specifically, when we extend $z_{max}$ to $\sim 2$, we find that 28,492 clusters between $1.5\lesssim z \lesssim 2$  can be additionally detected, $\sim 7\%$ more than before.
  The dark energy constraints obtained by extending the maximum redshift of the clusters to $z_{max} \sim 2$ are presented in Table~\ref{tab:w_0w_a_zmax2}. By extending the maximum redshift of the survey from $z_{max} \sim 1.5$ to $z_{max} \sim 2$, the dark energy constraints are tightened for both $w_0$CDM model and $w_0w_a$CDM model. The constraint on $w_0$ for $w_0$CDM model is improved by a factor of $\sim 1.06$ for the self-calibration scheme, and a factor of $\sim 1.28$ for the ``known SR" scheme. While for $w_0w_a$CDM model, the dark energy Figure of Merit is improved by a factor of $\sim 1.30$ for the self-calibration scheme, and a factor of $\sim 1.78$ for the ``known SR" scheme. As can be seen, even moderate detection of clusters at high redshift can tighten the constraints by an appreciable amount. The reason is that the behavior of dark energy deviate more from $\Lambda$CDM model in earlier Universe. Therefore if the survey can cover a large redshift range, a comparison of the behavior of dark energy at different redshifts helps to break parameter degeneracies. 
  
\begin{figure}
\begin{center}
{\includegraphics[angle=0, width=0.91\textwidth]{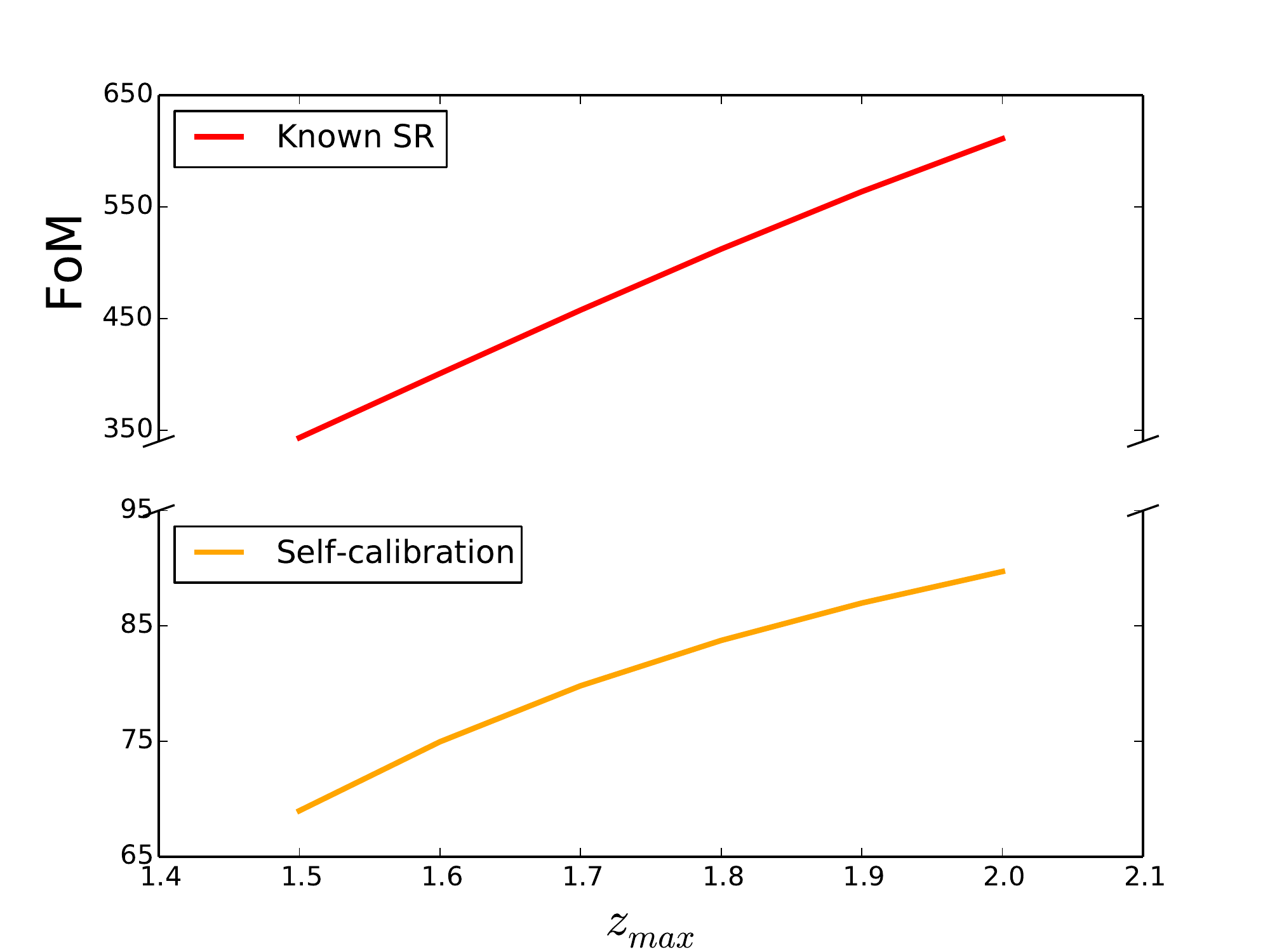}}
 \caption{\label{fig:FoM_maxz}The dark energy Figure of Merit (FoM) as a function of the maximum redshift $z_{max}$ of the CSST  clusters. Orange line stands for constraints obtained by self-calibration scheme. Red line stands for constraints obtained by ``known SR" scheme.}
\end{center}
\end{figure}

  In Figure~\ref{fig:FoM_maxz}, we show how the Figure of Merit for dark energy parameters in $w_0w_a$CDM model changes as the maximum redshift of the CSST clusters increases continuously. For both the self-calibration and ``known SR" schemes, the dark energy Figure of Merit increases steadily with $z_{max}$. Thus it is important to search for clusters at high redshift for stringent constraints on DE properties. It is also interesting to note that the Figure of Merit obtained from ``known SR" scheme increases more rapidly than that from self-calibration scheme. Thus high redshift clusters are more helpful to constrain dark energy if clusters have well calibrated scaling relation.
 
\begin{table}
\caption{\label{tab:w_0w_a_zmax2} Constraints on dark energy equation of state parameters by extending the maximum redshift of the CSST clusters to $z_{max} \sim 2$. Constraints shown are the marginalized 1$\sigma$ errors. The dark energy FoM is presented in the last row.}
\begin{center}
\begin{tabular}{c|c |c |c }
\hline\hline
Models & Parameters          &Self-calibration                                &Known SR\\
\hline\hline
$w_0$CDM           &$\Delta w_0$       &0.034                       &0.0094       \\
\hline
                           & $\Delta w_0$      &0.068                            &0.050         \\
$w_0w_a$CDM   &$\Delta w_a$       &0.32                              &0.17           \\
                 &FoM   &89.72           &  610.97                  \\
\hline
\end{tabular}
\end{center}
\end{table}

\subsection{The impact of redshift uncertainty}
\begin{figure}
\begin{center}
{\includegraphics[angle=0, width=0.91\textwidth]{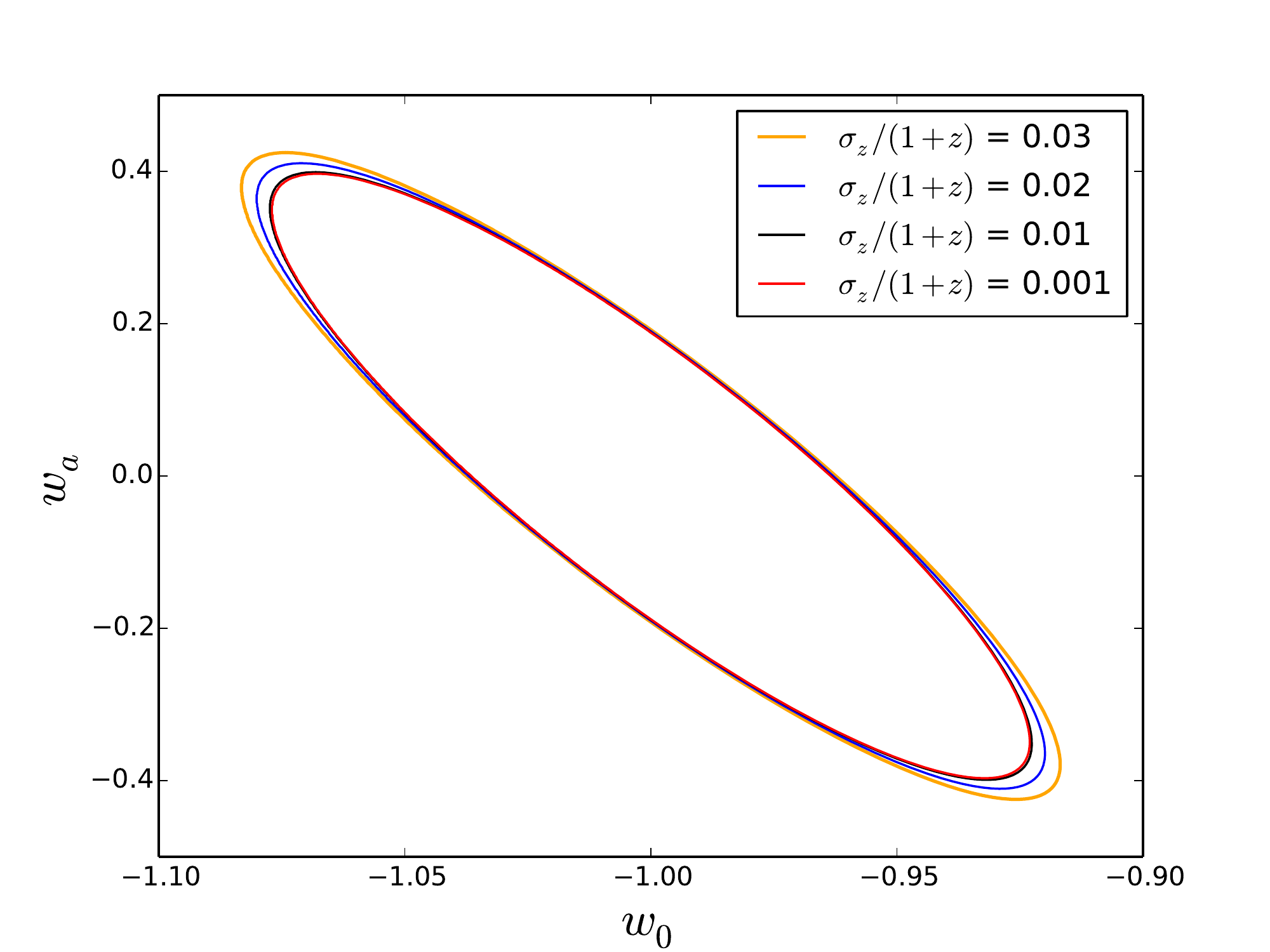}}
 \caption{\label{fig:ellipphotozRMpriorno}Impact on dark energy constraints from the CSST clusters'  redshift uncertainty. The contours are obtained with the self-calibration scheme. Orange, blue, black and red lines are for $\sigma_z/(1+z)$ equating to 0.03, 0.02, 0.01 and 0.001, respectively.}
\end{center}
\end{figure}

\begin{figure}
\begin{center}
{\includegraphics[angle=0, width=0.91\textwidth]{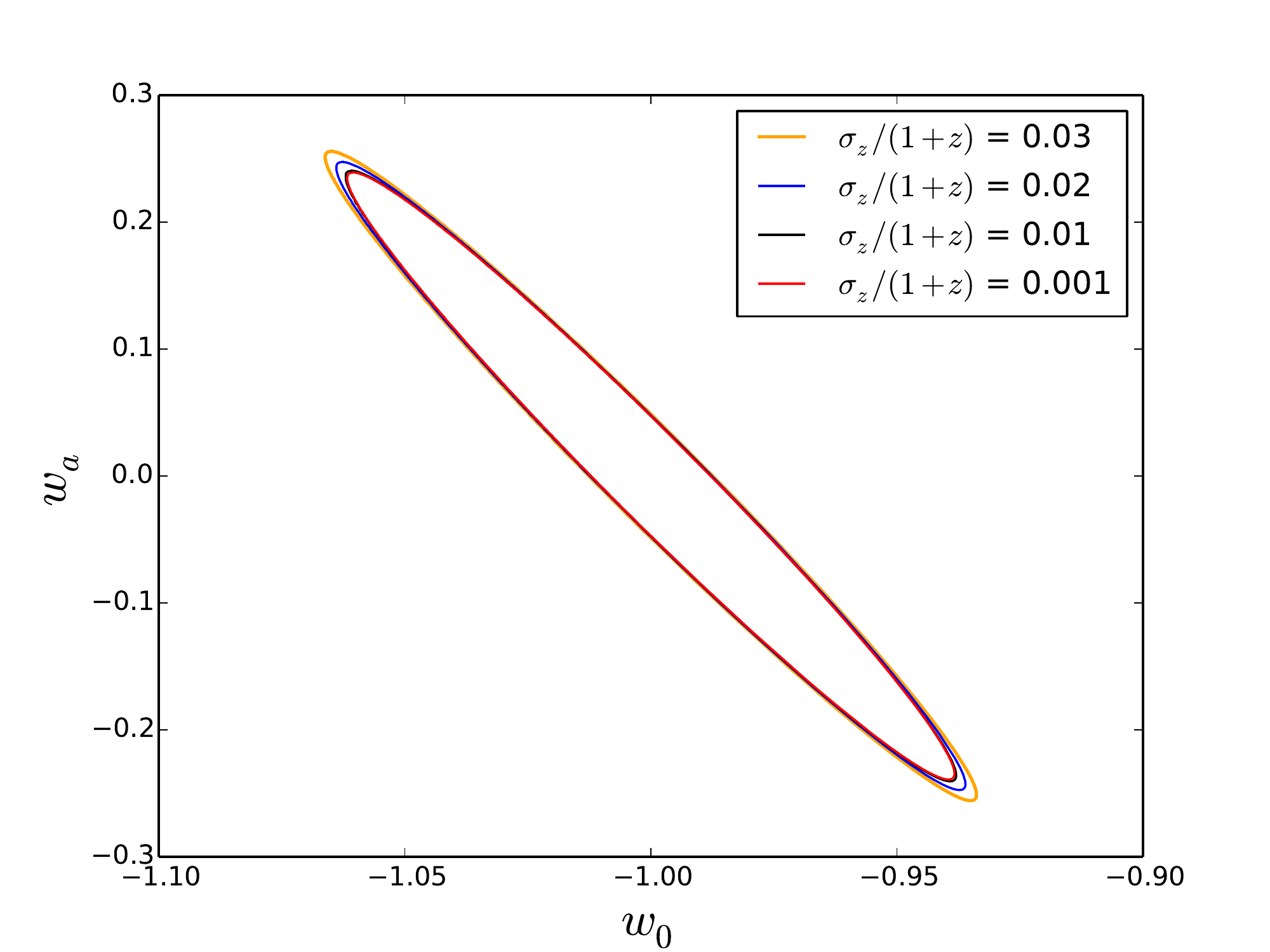}}
 \caption{\label{fig:ellipphotozRMfixed} Impact on the dark energy constraints from the CSST clusters'  redshift uncertainty. The contours are obtained with the ``known SR" scheme. Orange, blue, black and red lines are for $\sigma_z/(1+z)$ equating to 0.03, 0.02, 0.01 and 0.001, respectively.}
\end{center}
\end{figure}

The constraints above are obtained by assuming a somewhat optimistic redshift uncertainty of $\sigma_z/(1+z)$ = 0.001, which we expect to be achievable under the assumption that spectroscopic redshifts are available for all clusters of CSST. In this section, we study  the impact of less accurate redshifts of clusters on the cosmological constraints by assuming the CSST clusters have redshift accuracy of $\sigma_z/(1+z) =$ 0.03, 0.02, and 0.01, respectively. In Figure~\ref{fig:ellipphotozRMpriorno} and Figure~\ref{fig:ellipphotozRMfixed}, we show how the constraints on dark energy equation of state parameters change with respect to clusters' redshift accuracy. The error ellipses in Figure~\ref{fig:ellipphotozRMpriorno} are obtained with the self-calibration scheme, while those in Figure~\ref{fig:ellipphotozRMfixed} are obtained by ``known SR" scheme. In both figures, the DE constraints get tighter as CSST clusters' redshift uncertainty becomes smaller. However, the improvement in dark energy constraints is tiny from $\sigma_z/(1+z) = 0.01$ to $\sigma_z/(1+z) = 0.001$. For the self-calibration scheme, the DE FoM  improves from 68.44 to 68.99, while for ``known SR" scheme, the FoM improves from 340.09 to 343.25. In both cases, the improvement is less than 1 per cent. We conclude that the impact of CSST clusters' redshift uncertainty is negligible as long as the rms of redshift uncertainties is better than 0.01. According to \cite{2022MNRAS.513.3946W}, the redshift uncertainty of DES clusters is about 0.013 at redshifts $z \le 0.9$. Since CSST has two more wavebands than DES, we expect $\sigma_z/(1+z) = 0.01$ is achievable by CSST optical survey. We also find that if the clusters' redshift error degrades further to 0.03, FoM decreases only by a small amount of $\sim 9\%$ for both the self-calibration and ``known SR" schemes.

\subsection{Mass bias calibration}

In our analysis, the fiducial value of bias in observed mass is set to zero. However, if the observed cluster mass is biased, the derived dark energy parameter constraints will also be biased. It is interesting to ask that given the statistical accuracy achievable by the CSST clusters, what is the requirement for the calibration of bias in cluster mass. 

Consider data vector $D = \left\{ D_{\alpha} \right\}$ with covariance matrix C. The data bias $\Delta D$ induces bias in the i-th parameter $p_i$ as \citep{2010MNRAS.401.1399B}
\begin{equation}
\Delta p_i = \sum_{j}(F^{-1})_{ij}\sum_{\alpha,\beta} \frac{\partial \bar{D_{\alpha}}}{\partial p_j}(C^{-1})_{\alpha \beta} \Delta D_{\beta}.
\end{equation}

The bias calibration requirement is set by requiring the induced parameter bias to be smaller than the expected statistical variation in the cosmological parameters. The likelihood of the bias for a subset of interested parameters $\Delta p_A$ is determined by \citep{2010MNRAS.401.1399B}
\begin{equation}
\Delta \chi^2 = \Delta p_A^T F^{\prime} \Delta p_A,
\end{equation}
where $F^{\prime}$ is the marginalized Fisher matrix $ \left[ (F^{-1})_{AA}\right]^{-1}$.
In the case of two parameters,  the bias is smaller than the statistical error of 68 per cent when $\Delta \chi^2 < 2.3$.

The data elements $ D_{\alpha}$ in our analysis are the cluster number counts in mass and 
redshift bins. For simplicity we only compute how number counts be biased by mass bias parameter $B_{M,0}$, and we choose our parameters of interest as $\{w_0, w_a\}$. The requirement of $\Delta \chi^2 < 2.3$ translates to the requirement of $|B_{M,0}|< 0.30$. Thus the logarithm mass bias must be calibrated to 0.30 or better to avoid significant bias in the derived dark energy parameters. The condition $|B_{M,0}|< 0.30$ corresponds to $|\ln(\Mob/M)| < 0.30$, i.e., $0.74<\Mob/M < 1.35$, which is well satisfied by the weak lensing mass bias from a mock cluster survey with sky coverage of 5,000 $\rm deg^2$  and redshift range of $0.01< z < 1.51$ \citep{2020ApJ...891..139C}.

\subsection{Comparison with other optical cluster surveys}
\label{sec:comparison_de_cluster_csst}
    In this section, we make a comparison between our results of CSST and those of other optical cluster surveys, such as LSST and Euclid. Firstly, we compare our results for the CSST clusters to those of LSST clusters by \cite{Fang2007} utilizing a shear-selected cluster sample. The halo mass definition adopted by \cite{Fang2007} is based on identification of dark matter halos as spherical regions with a mean overdensity of 180 with respect to the background matter density at the time of identification. With a sky coverage of 18,000 $\rm deg^2$, they found that LSST can detect 276,794 clusters in the redshift range $0.1 \leq z \leq 1.4$, above the limiting halo mass of $\sim (0.6 - 4) \times 10^{14} M_{\odot}$. This number count is less than our result, since \cite{Fang2007} does not take into account uncertainties in the observable-mass scaling relation. Using cluster number counts alone, the forecasted Figure of Merit of dark energy equation of state parameters ($w_0,w_a$) is 14.1, weaker than our result. This is due to less cluster number count obtained by \cite{Fang2007}, and the WMAP priors adopted by them are much weaker than the  Planck priors we adopt.

We also compare our results for the CSST to those of Euclid by \cite{Sartoris2016}. The detection threshold of Euclid clusters is chosen such that $N_{500,c}/\sigma_\mathrm{field}$, the ratio between the number of cluster galaxies $N_{500,c}$ and the rms of field galaxies $\sigma_\mathrm{field}$, is greater than 3 (or 5).  The lowest limiting cluster mass for $N_{500,c}/\sigma_\mathrm{field} =3$ is $M_{200c} \sim 8 \times 10^{13} M_{\odot}$. With selection threshold $N_{500,c}/\sigma_\mathrm{field} = 5$, Euclid can detect $\sim 2 \times 10^5$ clusters up to redshift $z \sim 2$, with about $\sim 4 \times 10^4$ objects at $z \ge 1$. By lowering the detection threshold down to $N_{500,c}/\sigma_\mathrm{field} =3$, the total number of clusters rise up to $\sim 2 \times 10^6$, with $\sim 4 \times 10^5$ objects at $z\ge 1$. Our total number of clusters is between their estimated results for these two cases. Using cluster number counts alone, Euclid obtained dark energy FoM of $\sim 30$ for $N_{500,c}/\sigma_\mathrm{field} \ge 3$ in the self-calibration scheme. Though the abundance of Euclid clusters is greater than ours, the constraints we obtained are more competitive with theirs, since \cite{Sartoris2016} also includes curvature parameter $\Omega_k$ in their Fisher matrix analysis. We also note that the parametrization for the mass scatter in \cite{Sartoris2016} is different from ours. 

Finally, we notice that during the preparation of this paper, another result on dark energy constraints forecasted using the CSST galaxy clusters has appeared in \cite{Miao:2022hyp}. The cluster redshift range adopted by them is the same as ours. However, the limiting mass of clusters adopted by them ($M \geq 10^{14} h^{-1}M_{\odot}$) is higher than ours, and they do not take into account uncertainties in the observable-mass scaling relation, resulting in less clusters ($\sim 170,000$) than ours. They obtain the forecasted DE constraints of $\Delta w_0 = 0.13$ and $\Delta w_a = 0.46$ using the CSST cluster number counts, which are worse than ours, due to their much lower number of clusters and the fact that no Planck priors on the Hubble parameter and the cosmic baryon density are adopted in their analysis. 

\section{Conclusions}
\label{sec:conclusion_de_cluster_csst}
In this paper, we perform a comprehensive analysis of the constraints on dark energy for both constant ($w_0$CDM) and time-dependent ($w_0w_a$CDM) equation of state  expected from the CSST galaxy clusters. We make our forecast by adopting the Fisher matrix formalism tailored for measurements of cluster abundance. In the self-calibration scheme, we consider 14  parameters, seven of which characterize the cosmological model, while the remaining seven  model bias and  scatter in the scaling relation between the observed and true cluster masses for optically selected clusters. With the selection threshold in the observed halo mass of $M_{200m} \geq 0.836 \times 10^{14} h^{-1}M_{\odot}$, 414,669 clusters in the redshift range $0 \leq z \leq 1.5$ can be detected by the CSST, whose distribution  peaks at $ z \sim 0.6$. There are 103,069 clusters at $ z \geq 1.0$. The DE can be constrained to $\Delta w_0 = 0.036$ for $w_0$CDM model, $\Delta w_0 = 0.077$ and $\Delta w_a = 0.39$ for $w_0w_a$CDM model, with a  Figure of Merit of 68.99.

The self-calibration procedure would largely benefit from fixed  scaling relation. By fixing the seven nuisance parameters in our analysis, we get much tighter cosmological parameter constraints. We find that for the $w_0$CDM model, the constraint on $w_0$ is as good as $\Delta w_0 = 0.012$, an improvement by a factor of $\sim 3$ compared to self-calibration scheme. If $w_a$ is added as a free parameter, we obtain $\Delta w_0 = 0.062$ and $\Delta w_a = 0.24$ for the $w_0w_a$CDM model. The dark energy Figure of Merit  for  ($w_0,w_a$) is as high as 343.25, a great improvement by a factor of $\sim 5$ compared to the result of the self-calibration scheme. These results again highlight the importance of securing good knowledge of the observable-mass scaling relation. 

We investigate the possibility of tightening the dark energy constraints further by increasing the redshift extension of the CSST clusters. We extend the maximum redshift of the CSST clusters out to $z_{max} \sim 2$ and find an extra of 28,492 clusters between $1.5\lesssim z \lesssim 2$  can be detected. The dark energy Figure of Merit for  ($w_0,w_a$) increases to 89.72 and 610.97, for the self-calibration and ``known SR"  schemes, respectively, approximately improved by a factor of $\sim 1.30$ and $\sim 1.78$ from the results of $z_{max} \sim 1.5$. Thus, a small number of clusters at high redshift can tighten the cosmological constraints considerably, and high redshift clusters are more helpful to constrain dark energy with a better calibrated observable-mass scaling relation. 

We find that the impact of the redshift uncertainty of clusters on the constraints of dark energy is negligible as long as the  accuracy of redshift is better than 0.01, achievable by current DES survey. If the clusters' redshift error degrades further to 0.03, FoM decreases only by a small amount of $9\%$. We also find that the logarithm mass bias must be calibrated to $|B_{M,0}|< 0.30$ or better to avoid significant dark energy parameter bias.

In this work, we have focused on constraining dark energy parameters using cluster number counts alone. One can surely add in other cluster statistics to tighten the constraints with complementary information or better knowledge of systematics, for example the cluster power spectrum and the stacked lensing of clusters. On the other hand, various other fundamental problems can be investigated by using the CSST cluster sample, e.g., neutrino mass, primordial non-Gaussianity, modified gravity. We plan to investigate these prospects in future study.

\normalem
\begin{acknowledgements}
This work is supported by the National Key R$\&$D Program of China Grants No. 2022YFF0503404, 2021YFC2203102, by the National Natural Science Foundation of China Grants No. 12173036, 11773024, 11653002, 11421303 and 12073036, by the China Manned Space Project Grant No. CMS-CSST-2021-B01, by the Fundamental Research Funds for Central Universities Grants No. WK3440000004 and WK3440000005, and by the CAS Interdisciplinary Innovation Team.
\end{acknowledgements}
 
\appendix  
        
\section{Constraints on cosmological parameters}
      
\begin{figure}
\begin{center}
{\includegraphics[angle=0, width=0.91\textwidth]{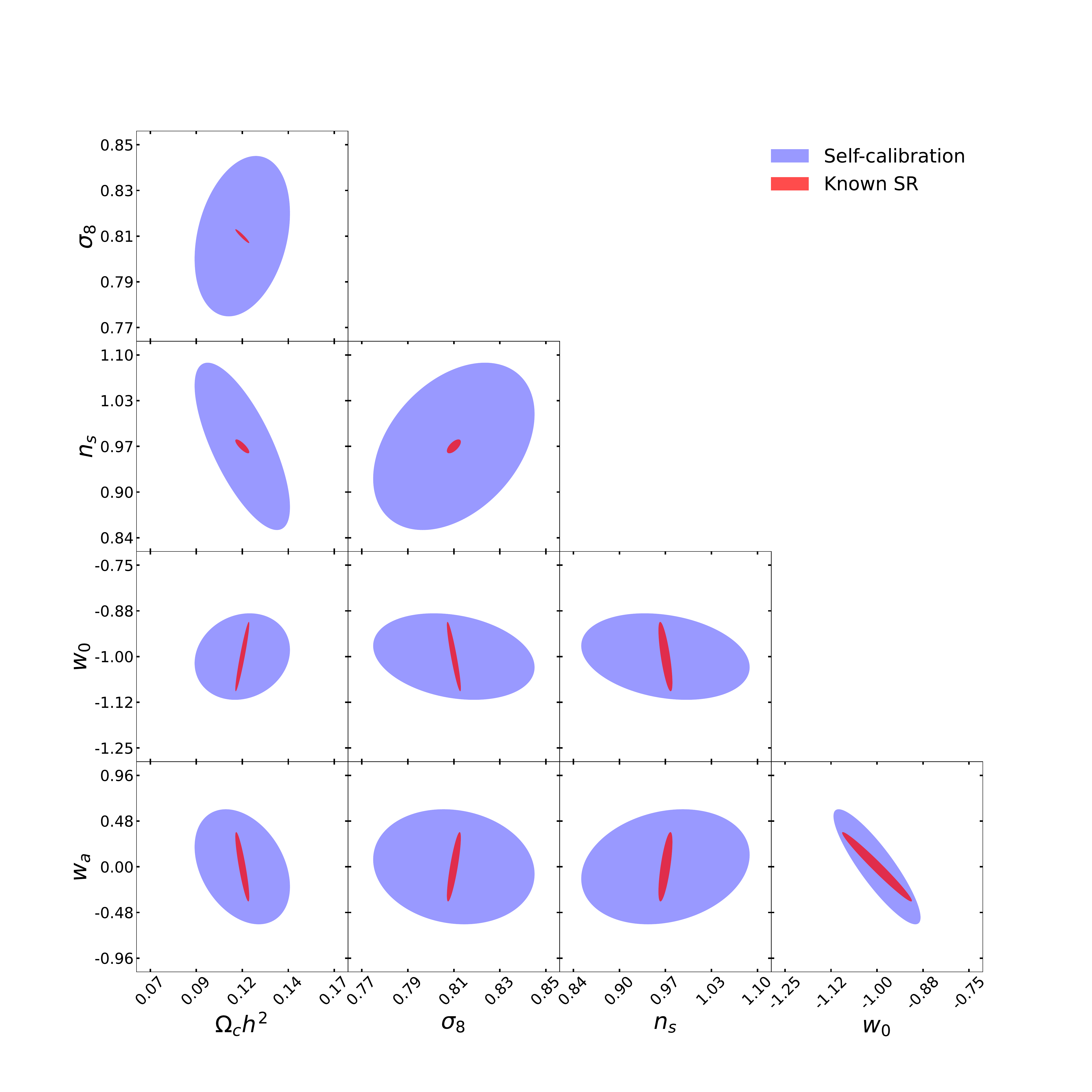}}
 \caption{\label{fig:ellipallparas}The contours of constraints (1 $\sigma$) on cosmological parameters for the $w_0w_a$CDM model obtained by the self-calibration scheme (blue) and ``known SR"  scheme (red), respectively.}
\end{center}
\end{figure}

In this Appendix, for completeness and comparison with other work, we display the constraints contours for all cosmological parameters for the $w_0w_a$CDM model, see Figure~\ref{fig:ellipallparas}. We do not show the contours for the Hubble parameter and the cosmic baryon density since we have used the Planck priors on these two parameters. The constraints are obtained by the self-calibration scheme (blue) and ``known SR"  scheme (red), respectively, assuming $z_{max} \sim 1.5$ and clusters' redshift uncertainty of 0.001. 
It is clear from these contours that the constraining power from CSST galaxy cluster survey will become much powerful if the scaling relation sector is better understood. The improvements from better knowledge of the observable-mass scaling relation are more pronounced for constraints on other cosmological parameters ($\Omega_ch^2$, $\sigma_8$ and $n_s$) than for $w_0$ and $w_a$. The degeneracy directions of some cosmological parameters are different for the two schemes since the inclusion of observable-mass scaling relation parameters in the self-calibration scheme will alter the degeneracy directions of the cosmological parameters in the  ``known SR"  scheme.

\bibliographystyle{raa}
\bibliography{DE_cluster_csst}

\end{document}